\documentclass[aps,
superscriptaddress,
prl,
amsmath,
footinbib,
amssymb,
bibnotes,
twocolumn
]{revtex4-2}

\usepackage{graphicx}
\usepackage{soul}      
\usepackage{color}
\usepackage{txfonts}
\usepackage[colorlinks, citecolor=blue,linkcolor=blue,urlcolor=blue]{hyperref}
\usepackage{verbatim}
\usepackage{multirow}
\usepackage{caption}
\usepackage{subcaption}
\graphicspath{{TmpFigs/}}

\usepackage{relsize}
\relscale{1.5}

 \usepackage{mathtools}
 \usepackage{wrapfig} 
 \usepackage{mathrsfs}
 \usepackage{amssymb} 
 \usepackage[x11names]{xcolor}
 \usepackage{amsbsy}
 \usepackage{marginnote}
 \usepackage{slashed}
 \usepackage{marvosym}
 \usepackage{pifont}
 \usepackage{mathbbol}
 \usepackage{wrapfig}
 \usepackage{upgreek}
 \usepackage{bbm}
 \usepackage{yfonts}
 \usepackage{xspace}
 \usepackage{lineno}
 \usepackage{tikz}
\usepackage[most]{tcolorbox}
 
 \newcommand{\bcen}{\begin{center}}
 \newcommand{\ecen}{\end{center}}
 \newcommand{\btab}{\begin{tabular}}
 \newcommand{\etab}{\end{tabular}}
 \newcommand{\bdes}{\begin{description}}
 \newcommand{\edes}{\end{description}}

 \newcommand{\beq}{\begin{equation}}
 \newcommand{\eeq}{\end{equation}}
 \newcommand{\bea}{\begin{eqnarray}}
 \newcommand{\eea}{\end{eqnarray}}

 \newcommand{\half}{\frac{1}{2}}
 \newcommand{\bary}{\begin{array}}
 \newcommand{\eary}{\end{array}}
 \newcommand{\benum}{\begin{enumerate}}
 \newcommand{\eenum}{\end{enumerate}}
 \newcommand{\bitem}{\begin{itemize}}
 \newcommand{\eitem}{\end{itemize}}

 \newcommand{\dou}{\partial}
 \newcommand{\leftjb} {[\![}
 \newcommand{\rightjb} {]\!]}
 \newcommand{\ju}[1]{ \leftjb #1 \rightjb }
 \newcommand{\D}[1]{\mbox{d}{#1}}

 \newcommand{\eqn}[1] {eqn.~(\ref{#1})}

 \newcommand{\fig}[1]{fig.~\ref{#1}}
 \newcommand{\Fig}[1]{Fig.~\ref{#1}}

 \newcommand{\calS}{\mathcal{S}}

 \newcommand{\ci}{\mathbbm{i}}

 \newcommand{\mylabel}[1]{\label{#1}}

\let\chapter\section
\let\section\subsection
\let\subsection\subsubsection

\newcommand{\oibook}[1]{}

\newcommand\figfac{0.9}

\newcommand{\IISc}{Centre for Condensed Matter Theory, Department of Physics, Indian Institute of Science, Bangalore 560012, India}
\newcommand{\MPIPKS}{Max-Planck-Institut f\"ur Physik komplexer Systeme, Dresden 01187, Germany}

\newcommand{\mytitle}{
Tearing Fractons}

\begin{document}

\title{\mytitle}

\author{Nandagopal Manoj}\email{nandagopalm@iisc.ac.in}\affiliation{\IISc}
\author{Roderich Moessner}\email{moessner@pks.mpg.de}\affiliation{\MPIPKS}
\author{Vijay B.~Shenoy}\email{shenoy@iisc.ac.in}\affiliation{\IISc}

\date{\today{}}
\begin{abstract} 
We offer a fractonic perspective on a familiar observation -- a flat sheet of paper can be folded only along a straight line if one wants to avoid the creation of additional creases or tears. Our core underlying technical result is the establishment of a duality between the theory of elastic plates and a fractonic gauge theory with a second rank symmetric electric field tensor, a scalar magnetic field, a vector charge and a symmetric tensor current. Bending moment and momentum of the plate are dual to the electric and magnetic fields, respectively. While the flexural waves correspond to the quadratically dispersing photon of the gauge theory, a fold defect is dual to its vector charge. Crucially, the fractonic condition constrains the latter  to move only along its direction, i.e.\ the fold's growth direction. By contrast, fracton
motion in the perpendicular direction amounts to tearing the paper. 
\end{abstract}

\pacs{}

\maketitle 




\noindent
{\it Introduction:}  Gauge theories play an eminent role in physics, all the way from  elementary particle theory at high energies to the emergent descriptions of topological quantum matter at low energies. The latter include instances such as the toric code \cite{Kitaev2003}, which has also played an important role in the context of topologically protected quantum computation. 



It is thus natural to look for generalisations of the gauge theories familiar from these settings. In this vein, fracton phases\cite{Chamon2005,Bravyi2011,Castelnovo2012,Castelnovo2012,Haah2011,Yoshida2013,Bravyi2013,Vijay2015,Vijay2016,Williamson2016,Hsieh2017} are perhaps the latest entrant. The most salient of their novel properties is the appearance of the eponymous fracton particles, which exhibit restricted, or `fractional' mobility, see e.g. the recent reviews \cite{Nandkishore2019,PretkoChenYou2020}).


It is those charges, but not their corresponding point dipoles, which are subject to fractonic mobility restrictions \cite{Pretko2017a,Pretko2017b}. 
In addition, these tensor gauge theories also support gapless ``photons'' much like the usual electromagnetism.  Such theories have also found generalizations to include extended fractons (line, surface like excitations) and varied dispersions of the gapless modes\cite{Pai2018,ShenoyMoessner2020}.

The demonstration\cite{Pretko2018} that the second rank tensor scalar charge theory is dual to the theory of crystalline elastic solids in two spatial dimensions firmly placed the physics of fractons in the realm of laboratory physics,
with fracton charges dual to immobile disclinations, while their dipoles are dual to dislocations which are subject to lesser mobility restrictions. The gauge structure, viewed from the perspective of the elastic solid, arises from the fact the dynamical equation of motion of the elastic solid written in terms of the stress tensor and the momentum density can be resolved by writing stress tensor and momentum density using symmetric tensor gauge fields (similar ideas are found in earlier literature, cf.~\cite{DietelKleinert2006} and references therein, although the notion of fractons was not introduced). These developments have led to related explorations\cite{Gromov2017,PretkoRadzihovsky2018,Pai2018,Kumar2019, Zhai2019,PretkoZhaiRadzihovsky2019, Gromov2019,Doshi2020}.

The present paper likewise identifies a  ``real life'' example of the fractonic gauge theory with {\em vector charges} introduced in refs.~\cite{Xu2006, RYX2016} (see also, \cite{Pretko2017a,Pretko2017b}). This gauge theory is formulated in terms of a symmetric second rank tensor electric field and a scalar magnetic field. We show that such a theory is dual to the theory of elastic plates\cite{LLElasticity,Mansfield1989}. 
Our duality, summarised in Table~\ref{tab:Duality}, maps the plate bending moment to the electric field of the gauge theory, the plate momentum density to the magnetic field, and the flexural wave in the plate to the photon of the gauge theory. Crucially,  ``fold defects'' of the plate are mapped to the fractonic vector charges of the gauge theory. This provides a fractonic perspective of the observation that a flat sheet of paper can be folded only along a straight line keeping the rest of the paper crease/tear free: the ``end point of a fold'' is a vector fractonic charge that can move only along the direction of the fold. Conversely, tearing of the paper can be understood as motion of a fractonic vector charge that violates the fractonic condition.
The gauge theory supports a photon which disperses as $\omega \textup{(frequency)} \sim |k|^2 \textup{(wavevector)}$. This corresponds to the flexural wave of the elastic plate which -- unlike e.g.\ the phonon -- has a quadratic dispersion.

\begin{table}[t!]
    \centering
    \begin{tabular}{||c||c||}
    \hline
    \hline
    Theory of Plates with Defects    &  Vector Charge Fracton Theory\\
    \hline \hline
    Bending moment ($M_{\alpha \beta}$)     &  Electric filed ($\epsilon_{\alpha \gamma} \epsilon_{\beta \delta} E_{\gamma \delta}$)\\
         \hline
    Momentum density ($P$) & Magnetic field ($B$) \\
    \hline
    Fold density ($\zeta_\alpha$) & Charge density $(\epsilon_{\beta \alpha} \rho_\beta)$ \\
    \hline
    Velocity curvature   & 
    \multirow{3}{1em}{}  \\
    $-$ Curvature velocity & Current density $(J_{\alpha \beta})$\\
     $ \left( \epsilon_{\alpha \delta} \epsilon_{\beta \gamma} \left( V^d_{\gamma \delta} - \dou_t R^d_{\gamma \delta} \right) \right)$ & \\
    \hline
    Flexural wave & Photon\\
    \hline \hline
    \end{tabular}
    \caption{Duality correspondence.}
    \label{tab:Duality}
\end{table}

The remainder of this paper provides the analysis underpinning these results. 
We have aimed to make the discussion largely self-contained, so that we repeat the central technical ingredients even when they are available elsewhere. We set the stage by  discussing the theories of vector charge fractons and elastic plates in turn, and then establish the duality mapping between the two. We conclude with a discussion.

\noindent
{\it Vector Charge Fracton Theory:} 
Consider a symmetric second rank electric field tensor $E_{\alpha \beta}$, and a scalar magnetic field $B$ in two dimensions, with position vector $x=(x_1,x_2)$, and time $t$.
Central to our discussion is a vector charge $\rho_\alpha(x,t)$ ($\alpha=1,2$) \cite{Xu2006, RYX2016,Pretko2017a,Pretko2017b,Seiberg2020}.  The form of Gauss' law which will endow the charge with fractonic character reads  
$\dou_\beta E_{\beta \alpha} = \rho_\alpha$
(for a more general version see  below) where repeated Greek (spatial) indices are summed over, and $\dou_\alpha$ is the derivative with respect to the spatial coordinate $x_\alpha$.

Electric and magnetic fields are encoded by a set of gauge fields $(\phi_\alpha, A_{\alpha\beta})$, with the vector field $\phi_\alpha(x,t)$ the analogue of the scalar potential  and the second rank tensor $A_{\alpha\beta}(x,t)$ that of the vector potential of Maxwell electromagnetism, via
\beq\label{eqn:EandB}
\begin{split}
E_{\alpha \beta} &= - \half (\dou_\alpha \phi_\beta + \dou_\beta \phi_\alpha) - \dou_t A_{\alpha \beta}, \;\;\;\; B = \epsilon_{\alpha \gamma} \epsilon_{\beta \delta} \dou_{\gamma} \dou_{\delta}  A_{\alpha \beta} , 
\end{split}
\eeq
invariant under the gauge transformation
induced by 
$f_\alpha(x,t)$:
\beq\label{eqn:GaugeTrans}
\begin{split}
\phi_\alpha  & \to \phi_\alpha + \dou_t f_\alpha, \;\;\; A_{\alpha \beta}  \to A_{\alpha \beta} - \half \left( \dou_\alpha f_\beta + \dou_\beta f_\alpha \right)\ .
\end{split}
\eeq
The complete theory is described by a Lagrangian density 
\beq\label{eqn:Lagrangian}
L = \half \kappa_{\alpha \beta \gamma \delta} E_{\alpha \beta} E_{\gamma \delta} - \frac{1}{2 \mu } B^2 - \rho_\alpha \phi_\alpha + J_{\alpha \beta} A_{\alpha \beta}
\eeq
with dielectric tensor $\kappa_{\alpha \beta \gamma \delta}$ and  magnetic permeability $\mu$, with $\kappa_{\alpha \beta \gamma \delta} = \kappa_{\beta \alpha \gamma \delta} = \kappa_{\alpha \beta \delta \gamma} = \kappa_{\gamma \delta \alpha \beta}$. $J_{\alpha \beta}$ is the second rank symmetric {\em current} tensor. Gauge invariance under \eqn{eqn:GaugeTrans} implies
\beq\label{eqn:Cont}
\dou_t \rho_\beta + \dou_\alpha J_{\alpha \beta} = 0\ .
\eeq 
The principle of least action provides two Maxwell equations
\begin{align}
\dou_{\alpha} \kappa_{\alpha \beta \gamma \delta} E_{\gamma \delta} & = \rho_\beta \label{eqn:Gauss}\\
\dou_t \kappa_{\alpha \beta \gamma \delta} E_{\gamma \delta} - \frac{1}{\mu} \epsilon_{\alpha \gamma} \epsilon_{\beta \delta} \dou_\gamma \dou_\delta B + J_{\alpha \beta} & = 0 \ . \label{eqn:Ampere}
\end{align}
From \eqn{eqn:EandB}, the vector charge version of Faraday law reads
\beq\label{eqn:Faraday}
\epsilon_{\alpha \gamma} \epsilon_{\beta \delta} \dou_{\gamma} \dou_{\delta}  E_{\alpha \beta} + \dou_t B = 0 \ .
\eeq

The fractonic character of the vector charge is revealed as follows. For a system with area $S$,  
\eqn{eqn:Cont}, implies conservation of the total charge $\int_S \D{^2x} \, \rho_\alpha$. Now, there is an additional conserved quantity, $\Sigma$, the moment of the vector charge 
\beq\label{eqn:Moment}
\begin{split}
\Sigma & = \int_S \D{^2 x} \, \epsilon_{\alpha \beta} x_{\alpha}  \rho_{\beta} \ .
\end{split}
\eeq
The consequence of this conservation law is that an isolated point charge -- the fracton -- {\it can only move along its own vector} $Q_\alpha$. This
can be illustrated by a point vector charge
$\rho_\alpha(x) = Q_\alpha \delta^{(2)}(x - x^0)$
located at $x^0$ ($\delta^{(2)}$ denotes the two dimensional Dirac delta function). The conservation law \eqn{eqn:Moment} imposes that any change of  $x^0$ must obey $x^0_\alpha \to x^0_\alpha + s {Q}_\alpha$ with $s\in \mathbb{R}$, to keep $\Sigma$ constant.

The theory supports a scalar, quadratically dispersing photon. For an isotropic system with ($\delta_{\alpha \beta}$ is the Kronecker delta)
\beq\label{eqn:kappa}
\kappa_{\alpha \beta\gamma \delta} = \kappa_1 \delta_{\alpha \beta} \delta_{\gamma \delta} + \frac{\kappa_2}{2} \left(\delta_{\alpha \gamma} \delta_{\beta \delta} + \delta_{\alpha \delta} \delta_{\beta \gamma} \right),
\eeq
$\omega(k) = \sqrt{\frac{(\kappa_1 + \kappa_2)}{\kappa_2 (2 \kappa_1 + \kappa_2) \mu}} |k|^2$ carrying  fields $B(x,t) = B_0 e^{\ci (k x - \omega(k) t)}$ and $E_{\alpha \beta}(x,t) = \frac{B_0}{\ci} \sqrt{\frac{(\kappa_1 + \kappa_2)}{\kappa_2 (2 \kappa_1 + \kappa_2) \mu}} \left[\delta_{\alpha \beta} -  \left(\frac{2 \kappa_1 + \kappa_2}{\kappa_1+\kappa_2} \right) \hat{k}_\alpha \hat{k}_\beta \right]  e^{\ci (k x - \omega(k) t)}$ (with $\ci = \sqrt{-1}$).

\begin{figure}
    \centerline{\includegraphics[width=\figfac\columnwidth]{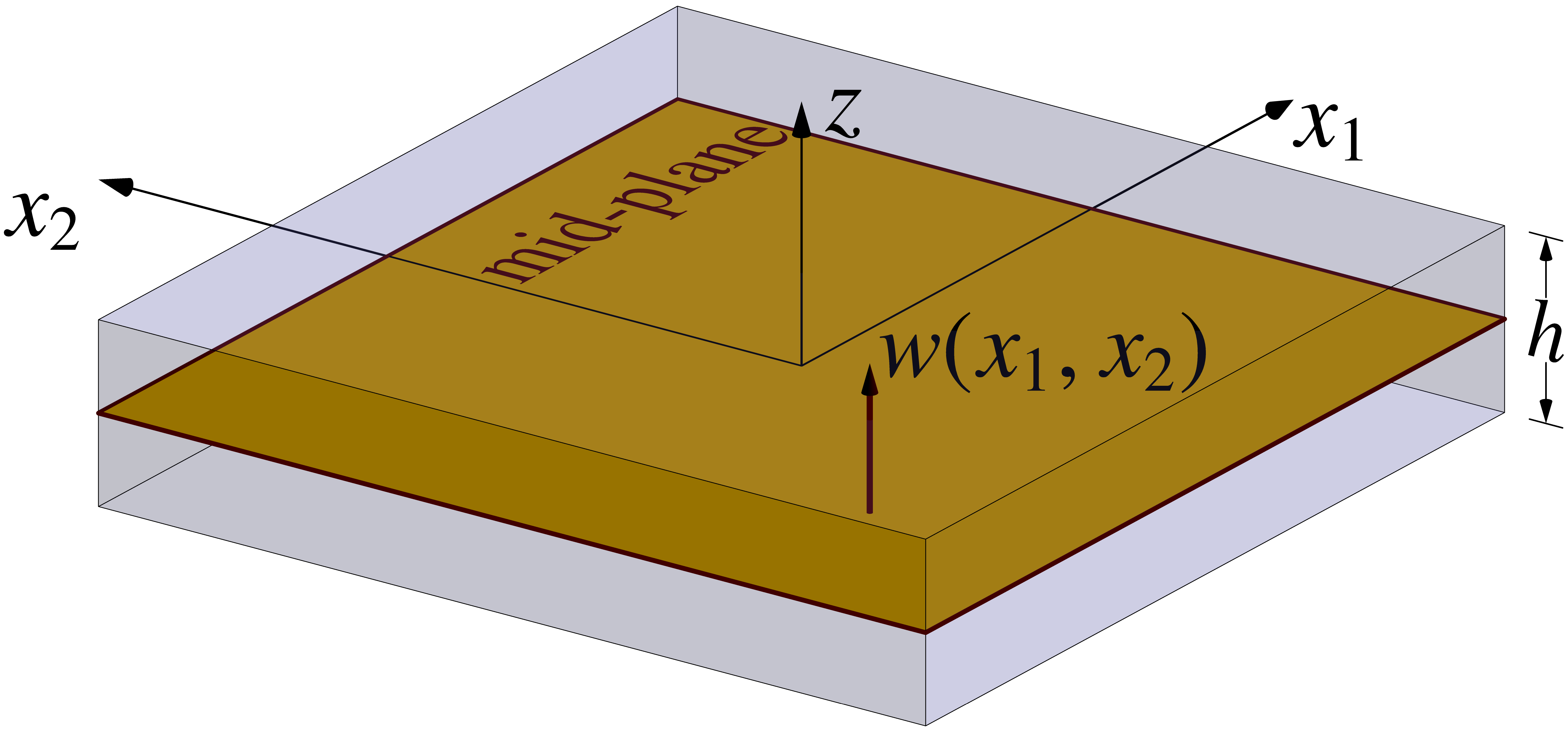}}
    \caption{Schematic of an elastic plate. The thickness of the plate is denoted by $h$. Lateral dimensions of the plate are much larger than the thickness. The deformation of the plate is described by the vertical (along the $z$-direction) displacement $w(x_1,x_2)$ of the mid-plane indicated. The displacement of any other material point is described by \eqn{eqn:ansatz}.}
    \label{fig:plate_scheme}
\end{figure}

\begin{figure*}
\centering
    \begin{subfigure}[b]{0.48\textwidth}
         \centering
         \includegraphics[width=\figfac\textwidth]{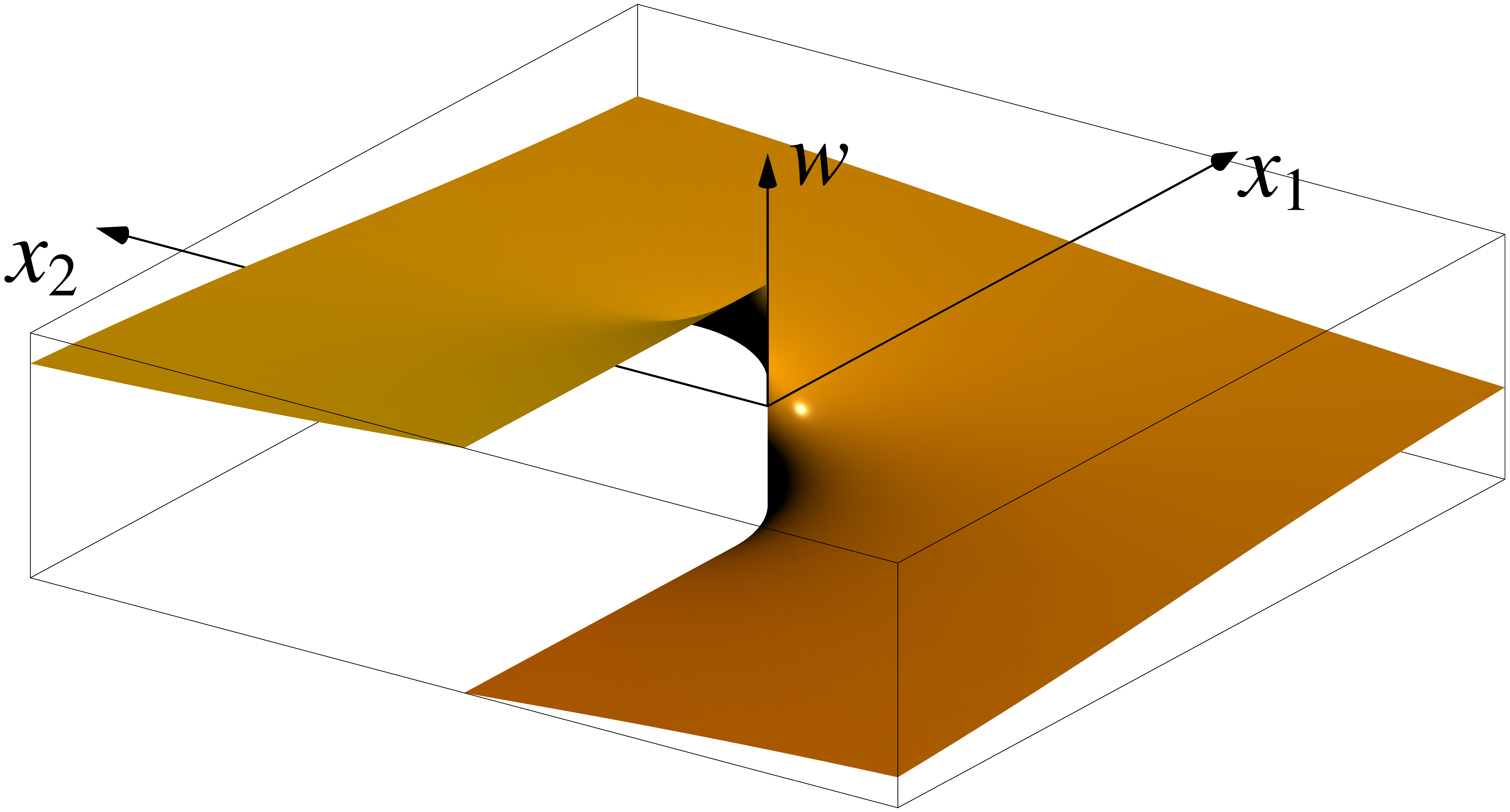}
         \caption{Tear defect}
         \label{fig:puretear}
     \end{subfigure}
     \hfill
     \begin{subfigure}[b]{0.48\textwidth}
         \centering
         \includegraphics[width=\figfac\textwidth]{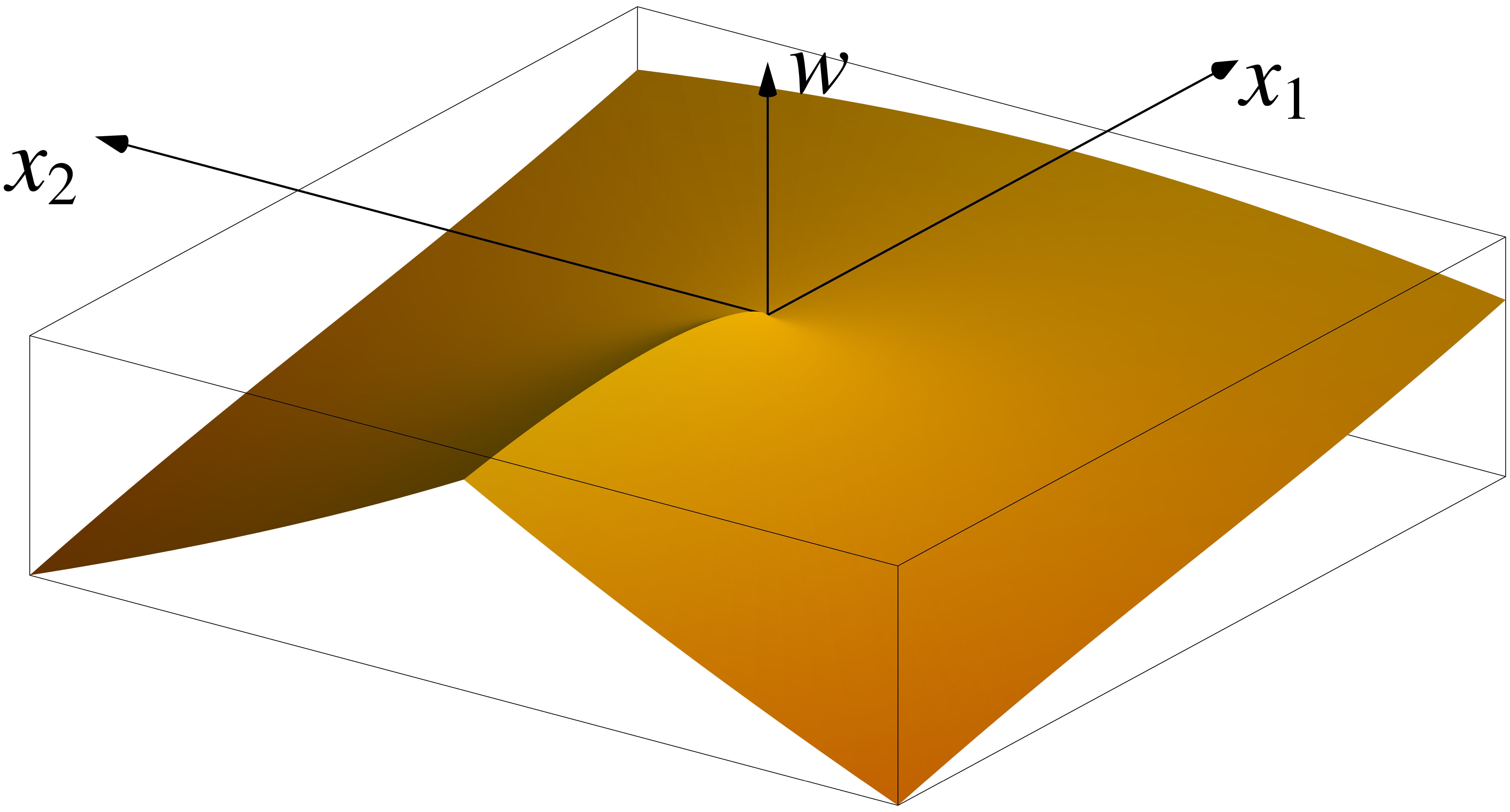}
         \caption{Fold defect}
         \label{fig:purefold}
     \end{subfigure}
     \caption{{\bf (a)}
     Tear defect located at the origin with $w(x_1,w_2) = \frac{1}{2\pi} \arctan{\frac{x_2}{x_1}}$ and tear defect density (\eqn{eqn:tauTheta}) $\tau = \delta^{(2)}(x)$ that satisfies \eqn{eqn:teardefect} with $a=1$.
     {\bf (b)} Fold defect located at the origin with $w(x_1,x_2) = \frac{1}{2 \pi} \left[ x_1 \ln(|x|) - x_2 \arctan{\left( \frac{x_2}{x_1} \right)} \right] $ and defect density $\zeta_\alpha = \delta_{\alpha 2} \delta^{(2)}(x)$ that satisfies \eqn{eqn:folddefect} with $\psi_{\alpha} = \delta_{\alpha 2}$. 
      For the tear defect {\bf (a)} the slopes $\theta_\alpha = \dou_\alpha w$ and curvatures $R_{\alpha \beta} = \dou_\alpha \dou_\beta w$ are smooth everywhere (except the origin), while for  fold defect ${\bf (b)}$ only the latter applies. For the fold defect the displacement is continuous and single valued. We chose the branch of $\arctan$ that returns a value in $[-\pi,\pi]$, such that $\lim_{\varepsilon \to 0}\arctan\frac{-\epsilon}{-1} = -\pi$.
     }
    \label{fig:alldefects}
\end{figure*}

\noindent
{\it Elastic Plates:} The physics of a three dimensional elastic solid\cite{LLElasticity} is described by a displacement field $u_i(r,t)$, a symmetric strain tensor field $\epsilon_{ij}(r,t)$~and a symmetric stress tensor field $\sigma_{ij}(r,t)$, where $r$ is position vector of a material particle in three dimensions (latin indices  $i,j$ etc., run over all spatial coordinates, $i=1,2,3$). The strain tensor $\epsilon_{ij} = \half \left( \dou_i u_j + \dou_j u_i\right)$ and the stress tensor are related by an elastic constitutive relation $\sigma_{ij} = C_{ijkl} \epsilon_{kl}$ where the elastic tensor $C_{ijkl}$ obeys $C_{ijkl} = C_{jikl} = C_{ijlk} = C_{klij}$. For isotropic solids, which is our focus here, $C_{ijkl} = \frac{E}{1+\nu} \left[\frac{\nu }{(1 - 2 \nu)} \delta_{ij}\delta_{kl} + \half \left( \delta_{ik} \delta_{jl} + \delta_{il} \delta_{jk} \right)\right] $ where $E$ and $\nu$ are, respectively,  Young's modulus and Poisson's ratio. The Lagrangian density of the system is given by
\beq\label{eqn:Elast3D}
L = \half \lambda \dou_tu_i  \dou_tu_i - \half C_{ijkl} \epsilon_{ij} \epsilon_{kl}
\eeq
with $\lambda$  the mass density of the solid.
For a  ``slender'' body (see the plate in  \Fig{fig:plate_scheme}) with thickness $h$ much smaller than its lateral dimensions, $L_\alpha$, an effective low energy theory -- plate theory -- that becomes increasing accurate as $h/L_\alpha \to 0$ can be developed. This ansatz uses a coordinate system  $(r_1,r_2,r_3) = (x_1,x_2,z) \equiv (x,z)$ where $z$ is the coordinate normal to the plane (henceforth called the vertical). It uses the {\em mid-plane} displacement normal to itself, denoted by $w(x) = u_3(x,0)$ to encode the {\it full} dynamics of the three-dimensional plate via  \cite{LLElasticity, Mansfield1989}:
\beq\mylabel{eqn:ansatz}
u_1(x,z) = - z \dou_1 w, \;\;\; u_2(x,z) = -z \dou_2 w
\eeq
In addition, the stress component $\sigma_{33}$ vanishes everywhere. With these given, we have $\sigma_{\alpha \beta} = \bar{C}_{\alpha \beta \gamma \delta} \epsilon_{\alpha \beta}$, where $\bar{C}_{\alpha \beta \gamma \delta} = \frac{E}{1-\nu^2} \left[(1-\nu) \delta_{\alpha \beta} \delta_{\gamma \delta} + \frac{\nu}{2}  \left( \delta_{\alpha \gamma} \delta_{\beta \delta} + \delta_{\alpha \delta} \delta_{\beta \gamma}  
\right) \right]$. The effective Lagrangian density {\em on the mid-plane} reads
\beq\label{eqn:PlateLagrangian}
L = \frac{\lambda_p}{2} \dou_t w \dou_t w - \half D_{\alpha \beta \gamma \delta} (\dou_{\alpha} \dou_{\beta} w) (\dou_{\gamma} \dou_\delta w),
\eeq
where $\lambda_p = \lambda h$
is the mass per unit area of the plate, and 
\beq\label{eqn:BModTensor}
D_{\alpha \beta \gamma \delta} = D \left[\nu \delta_{\alpha \beta} \delta_{\gamma \delta} + \frac{(1-\nu)}{2} \left( \delta_{\alpha \gamma} \delta_{\beta \delta} + \delta_{\alpha \delta} \delta_{\beta \gamma} \right) \right], 
\eeq
the bending modulus tensor, with
$D = {E h^3}/({12 (1-\nu^2)})$
the bending modulus. The quantity $\dou_\alpha \dou_\beta w (\equiv R_{\alpha \beta})$ is the curvature tensor induced by the deformation $w$. The bending modulus obtains the bending moment from the curvature via $M_{\alpha \beta} = D_{\alpha \beta \gamma \delta} \dou_{\gamma} \dou_{\delta} w$. The plate supports ``flexural waves'', $w(x,t) = w_0 e^{\ci (k x - \omega(k) t)}$ which disperse as $\omega(k) = \sqrt{\frac{D}{\lambda_p}} |k|^2$. 

The usual discussion of the theory of plates considers smooth single valued displacement fields $w(x,t)$. Here we discuss the types of defects that arise in theory of plates. For this, we introduce an additional quantity, $\theta_\alpha = \dou_\alpha w$,  the ``slope of the deformed mid-plane'' (see \fig{fig:plate_scheme}), or equivalently the rotation of material fibre vertical to the mid-plane. 

\noindent
{\em Tear Defect:} 
First, a tear defect of strength $a$ located at a point $x^0$, implies for a closed contour $C$ (which avoids the point $x^0$),
\beq\label{eqn:teardefect}
\oint_C \D{x_\alpha} \theta_\alpha = \begin{cases}
a \,\, \mbox{if $C$ encloses $x^0$} \\
0 \,\, \mbox{if $C$ does not enclose $x^0$}.
\end{cases}
\eeq
\Fig{fig:puretear} shows an example of an isolated tear defect. This illustrates the multi-valued nature of the displacement field $w$ resulting in the tear of the plate. The multivalued $w$ field is such that  both the slope field and the curvature fields are smooth everywhere except at the location of the defect where they diverge.
We define a density of such tear defects
\beq\label{eqn:tearjump}
\int_A \D{^2 x} \, \tau = \oint_C \D{x_\alpha} \theta_\alpha = \ju{w}_C
\eeq
with $A$  the area enclosed by $C$, and $\ju{w}_C$ the jump in the displacement field obtained upon traversing the contour $C$ and
\beq\label{eqn:tauTheta}
\tau = \epsilon_{\alpha \beta} \dou_\alpha \theta_{\beta}.
\eeq
For the single tear defect located at $x^0$ discussed above $\tau(x) = a \delta^{(2)}(x-x^0)$ (see \fig{fig:puretear} where $a=1$ and $x_0$ is the origin).  



\noindent
{\em Fold Defect:} Second, a fold defect (\fig{fig:purefold}) of strength $\psi_\alpha$ located  at $x^0$ produces
\beq\label{eqn:folddefect}
\oint_C \D{x}_\alpha \, R_{\alpha \beta} = \begin{cases}
\psi_\beta \,\, \mbox{if $C$ encloses $x^0$} \\
0 \,\, \mbox{if $C$ does not enclose $x^0$}
\end{cases}
\eeq
Its curvature tensor is smooth everywhere except at the location of the defect. Further, the defect goes along with a fold in the plate, which terminates at the defect. For a continuous distribution of fold defects, the fold density $\zeta_\alpha$  is given by
\beq\label{eqn:foldjump}
\oint_C \D{x}_\alpha R_{\alpha \beta} = \int_A \D{^2 x} \zeta_\alpha = \ju{\theta_\alpha}_C 
\eeq
where $\ju{\theta_\alpha}_C $ is the net jump in the slope when traversing the closed contour $C$ that encloses the area $A$. We thus get
\beq\label{eqn:zetaR}
\zeta_\alpha = \epsilon_{\beta \gamma} \dou_\beta R_{\gamma \alpha}\ .
 \eeq
Again, for the fold defect shown in \fig{fig:purefold}, $\zeta_\alpha = \delta_{\alpha 2} \delta^{(2)}(x)$.

An important point to be noted is that for the given defect configuration, there are many distinct displacement fields $w^d(x)$ that will satisfy conditions of the type \eqn{eqn:tearjump} and \eqn{eqn:foldjump}. For example, the tear defect $\tau = a \delta^{(2)}(x)$ is equally well described by a  different displacement field than that shown in \fig{fig:puretear} by  a different choice of the branch of the $\arctan$ function. This situation is akin to the description of a superfluid vortex, or a vector potential of a magnetic monopole, and suggests the presence of an underlying gauge structure in the theory when defects are present \cite{Kleinert1989}.

We now write the the displacement $w$ as a sum of two terms
\beq\label{eqn:wSplusD}
w(x,t) = w^s(x,t) + w^d(x,t)
\eeq
where $w^s(x,t)$ is the smooth/single-valued part of the deformation and $w^d(x,t)$ is a multi-valued field that describes the defects, and insert this into the Lagrangian density (\eqn{eqn:PlateLagrangian}). %


\noindent
{\it Duality:} To establish the duality between plate and fracton theories, we introduce two Hubbard-Stratonovich fields $P(x,t)$ (momentum density)  and $M_{\alpha \beta}(x,t)$ (bending moment) to treat the kinetic and potential energy terms in the Lagrangian \eqn{eqn:PlateLagrangian}:
\beq\label{eqn:HS}
\begin{split}
\calS[w,P,M_{\alpha \beta}]  = \int \D{t} \D{^2 x} & \left[\half D^{-1}_{\alpha \beta \gamma \delta} M_{\alpha \beta} M_{\gamma \delta} - \frac{1}{2 \lambda_p} P^2  \right. \\
  + P (\dou_t w^s + \dou_t w^d) & - M_{\alpha \beta} (\dou_{\alpha} \dou_{\beta}w^s + \dou_{\alpha} \dou_{\beta}w^d) \Big]
\end{split}
\eeq
Integrating out the smooth displacement field $w^s$ gives
\beq\label{eqn:Equil}
\dou_t P + \dou_\alpha \dou_\beta M_{\alpha \beta} = 0.
\eeq
We can now identify
\beq\label{eqn:BPEM}
\begin{split}
B(x,t) \equiv P(x,t), \;\;\;\;\epsilon_{\alpha \gamma} \epsilon_{\beta \delta} E_{\gamma \delta}(x,t)
\equiv  M_{\alpha \beta}(x,t) \ .
\end{split}
\eeq
This shows that \eqn{eqn:Equil} can be resolved (identically satisfied) if $B$ and $E_{\alpha \beta}$ are expressed through the gauge fields $\phi_\alpha$ and $A_{\alpha \beta}$ as in \eqn{eqn:EandB}. Using \eqn{eqn:BPEM} and \eqn{eqn:EandB}, we the action \eqn{eqn:HS} in terms of the gauge fields reads
\beq\label{eqn:Dual}
\begin{split}
\calS[w^d,\phi_\alpha,A_{\alpha \beta}]  = \int \D{t} \, \D{^2 x} & \left[\half \kappa_{\alpha \beta \gamma \delta} E_{\alpha \beta} E_{\gamma \delta} - \frac{1}{2 \mu} B^2  \right. \\
\!\!\!\!\!\!+  (\epsilon_{\alpha \delta} \epsilon_{\beta \gamma} \dou_\alpha \dou_\beta A_{\gamma \delta}) \dou_t w^d 
+  \epsilon_{\alpha \gamma} \epsilon_{\beta \delta} & \left.  \left( \half \left(\dou_\gamma \phi_{\delta} + \dou_\delta \phi_\gamma \right) + \dou_t A_{\gamma \delta} \right) \dou_\alpha \dou_\beta w^d \right]
\end{split}
\eeq
where, using \eqn{eqn:kappa}, the following identifications are made 
$\kappa_{\alpha \beta \gamma \delta} \equiv D^{-1}_{\alpha \beta \gamma \delta}, 
\kappa_1 \equiv -\frac{\nu}{D ( 1 - \nu^2)}, 
\kappa_2 \equiv \frac{1}{D (1-\nu)},
\mu \equiv \lambda_p$
Finally, the action \eqn{eqn:Dual}, after suitable integration by parts of the last two terms,  reduces exactly to the action governed by the Lagrangian density \eqn{eqn:Lagrangian} of the vector charge fracton theory. The dual charges and currents of the fracton theory are
\begin{align}\label{eqn:rhoJ}
\rho_\alpha & \equiv \epsilon_{\alpha \delta} \epsilon_{\beta \gamma} \dou_\beta R^d_{\gamma \delta} = \epsilon_{\alpha \delta}  \zeta_\delta \\
J_{\alpha \beta} & = \epsilon_{\alpha \delta} \epsilon_{\beta \gamma} \left( V^d_{\gamma \delta} - \dou_t R^d_{\gamma \delta} \right)
\end{align}
where we have used \eqn{eqn:zetaR}, and 
\beq
R^d_{\alpha \beta} = \dou_\alpha \dou_\beta w^d, \;\; V^d_{\alpha \beta} = \dou_\alpha \dou_\beta (\dou_t w^d)
\eeq
are respectively the defect curvature field and defect ``velocity curvature'' field. The duality is summarized in table~\ref{tab:Duality}. We note here the duality of the vector charge gauge theory to a scalar field was noted in ref.~\cite{Xu2006}, although the connection to the theory of plates was not discussed.

\noindent
{\it Discussion:} We first address
the connection between the vector charges of the gauge theory and the fold defects of plates. Consider a fold on a plate that lies long the $x_1$-axis terminating at the origin (\fig{fig:purefold}), such that the normals to the plate just below and just above the negative $x_1$ axis are tilted by a small angle $\psi$, corresponding to $\zeta_\alpha = \psi \delta_{\alpha 2}  \delta^{(2)}(x)$
indicating a jump in the $2$-component of the slope of the plate when traversing across the negative $x_1$-axis of $\psi$. 
This  corresponds to a fractonic vector charge, \eqn{eqn:rhoJ}, of
$\rho_\alpha = \psi \delta_{\alpha 1} \delta^{(2)}(x)$
i.~e., a ``charge along the $x_1$ direction'' located at the origin. Following the discussion near \eqn{eqn:Moment} we see that this charge is allowed to move only along the $x_1$ axis. Viewed, again, from the perspective of plates, we see that a point fold defect can only move -- and the fold extended -- in a direction perpendicular to its strength (which is a vector).
We thus obtain a fractonic perspective on the observation that a flat sheet of paper can be folded smoothly (without the creation of additional creases/tears) only along a straight line. 

\begin{figure}
\centerline{\includegraphics[width=\figfac\columnwidth]{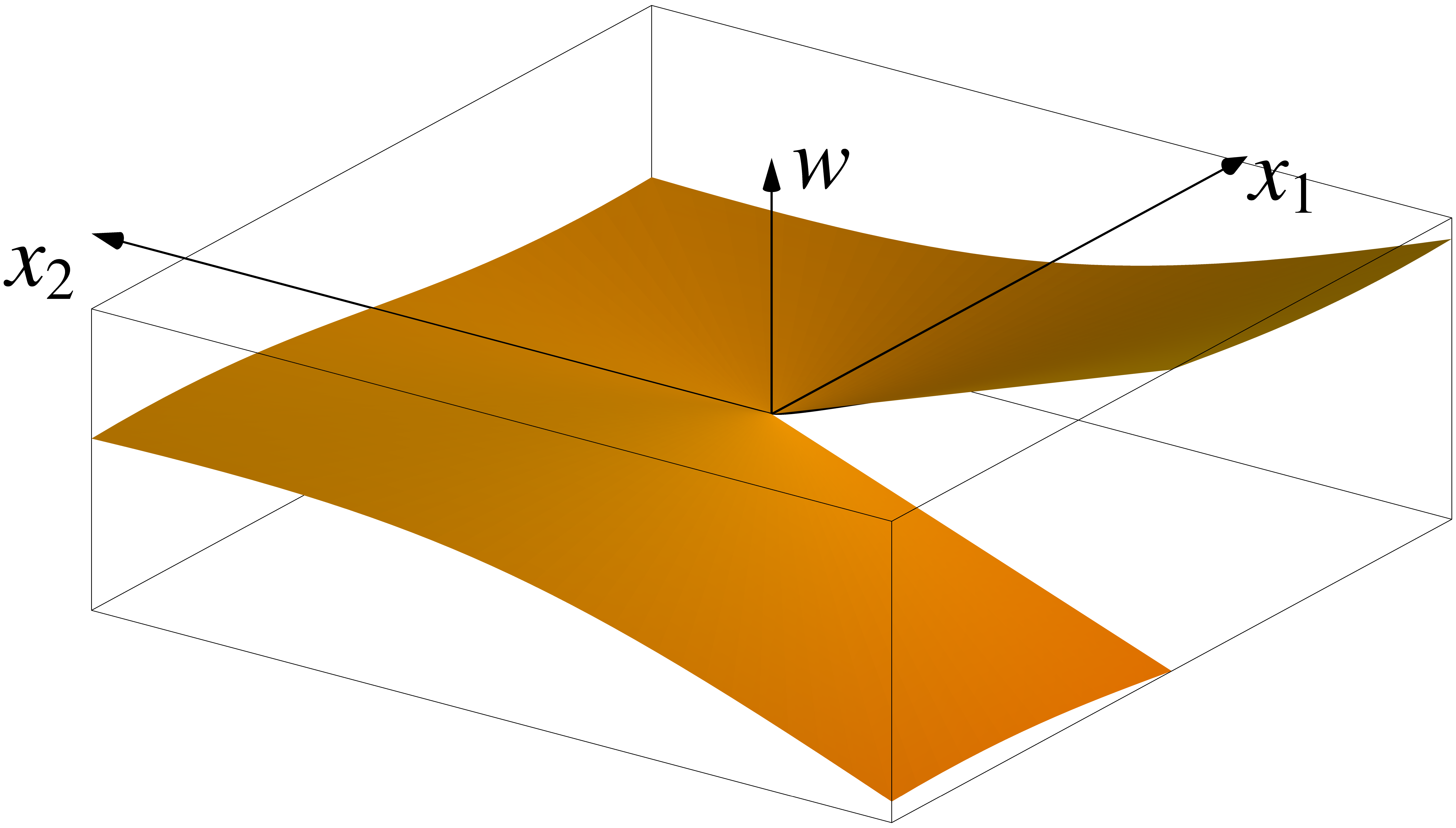}}
\caption{ Configuration obtained by transporting a fold defect with strength $\psi_\alpha = \delta_{\alpha 2}$ from $x_2 = -\infty$ along the $x_2$ axis to the origin.
The displacement field $w(x_1,x_2) = -\frac{1}{2 \pi} \left[ x_2 \ln(|x|) + x_2 \arctan{\left( \frac{x_1}{x_2} \right)} \right] $ satisfies \eqn{eqn:folddefect} and its curvature fields are identical to that in \fig{fig:purefold}.  }
\label{fig:violation}
\end{figure}

Then, what about the tear and fractons? The connection can be seen by noting two points:
First, note that the tear defect, \eqn{eqn:tauTheta} in fact is a {\it dipole} of fold defects. For instance, the tear defect $\tau(x) = a \delta^{(2)}(x)$ arises from a fold pair with $\zeta_\alpha (x) = \psi \delta_{\alpha 2} \left( \delta^{(2)}(x + \varepsilon \hat{e}_2) - \delta^{(2)}(x) \right)$, where $\hat{e}_2$ is the unit vector along the $x_2$ direction, such that $\lim_{\varepsilon \to 0} \psi \epsilon = a$. More generally, a dipole of fold defects located at a point $x^0$ has 
$ \zeta_\alpha(x) = \lim_{\varepsilon \to 0} \psi_\alpha \varepsilon \hat{n}_\beta \dou_\beta \delta^{(2)}(x - x^0) = a_\alpha \hat{n}_\beta \dou_\beta\delta^{(2)}(x - x^0)$,
where $\hat{n}_\beta$ is a unit vector. This results in $(\lim_{\varepsilon \to 0} \psi_\beta \varepsilon = a_\beta)$, $
\rho_\alpha (x) = \epsilon_{\alpha \beta} a_\beta \hat{n}_\gamma \dou_\gamma \delta^{(2)}(x-x^0)$
which is a tear defect of strength of $a_\gamma \hat{n}_\gamma$. From \eqn{eqn:Moment}, 
$\Sigma_{\textup{dipole}} = a_\gamma \hat{n}_\gamma \approx \varepsilon \psi_\alpha \hat{n}_\gamma$, independent of $x^0$.

Second, consider the displacement of a defect $\zeta_\alpha = \psi_{\alpha}\delta^{(2)}(x - x_0)$ from $x_0$ to $x_0 + \varepsilon \hat{n}$ along the unit vector $\hat{n}$. Using \eqn{eqn:Moment} we see that the moment $\Sigma$ changes by an amount $\varepsilon \psi_\alpha \hat{n}_\alpha$. The moment $\Sigma$ can be conserved by creating an additional defect whose moment is $-\varepsilon \psi_\alpha \hat{n}_\alpha$, which we see immediately is a dipole or tear defect! 

A fold defect thus sheds tear defects to compensate its  moment change when displaced in the `wrong' direction.
\Fig{fig:violation}  shows the configuration of the plate obtained  by transporting a fold defect with strength $\delta_{\alpha 2}$ from $x_2 = -\infty$ along the $x_2$ axis to the origin, shedding a uniform density of tear defects along its path: the fracton tears the plate in the process!  
(Note that this configuration 
satisfies \eqn{eqn:folddefect} precisely as the fold defect shown in \fig{fig:purefold}: both defect configurations have the same curvature distributions.)


We hasten to point out that, while the duality  provides insights into commonly observed phenomena related to folding and tearing, it does not take into account the `nonlinear' irreversible/plastic processes that accompany the motion of defects (treatment similar to \cite{PretkoRadzihovsky2018}): these are not included in the dual  fracton theory. 

Indeed, the crumpling of paper\cite{Cerda1998,Witten2007, Gottesman2018},  governed by the interplay of out of plane deformation (described by the field $w$) and in-plane stretching, is not considered at all in the linear theory presented here. The structures that arise in crumpled paper attempt to minimize the in-plane stretching energy  maximizing regions where Gaussian curvature vanishes. 
It will be interesting to explore generalizations of our formulation to situations where in plane deformation and out of plane deformation are coupled (Fl\"oppl-von Karman theory\cite{LLElasticity,Mansfield1989}).

We conclude by noting that our work provides an example of a physical system embodying fractonic physics adding to a growing list \cite{PretkoRadzihovsky2018,Benton2016,You2019,Taylor_2020,Khemani_2020,Sous2020}. It will be interesting to explore other systems to find further examples. The general structure of fracton theories discussed in ref.~\cite{ShenoyMoessner2020} might provide clues to look for the dual physical realizations.

\noindent
{\it Acknowledgements:} NM acknowledges the KVPY programme, and VBS thanks SERB, DST for support. This work was in part supported by the Deutsche Forschungsgemeinschaft  under SFB 1143 (project-id 247310070) and  cluster of excellence ct.qmat (EXC 2147, project-id 390858490).

\bibliographystyle{apsrev4-2}
\bibliography{fracton}
\end{document}